\pdfoutput=1
\documentclass[11pt]{article}

\usepackage{amsmath}
\usepackage{amsfonts}
\usepackage{amssymb}

\usepackage{booktabs} 

\usepackage{todonotes}
\usepackage{lineno}
\usepackage{url}
\usepackage[algo2e,ruled,linesnumbered,vlined]{algorithm2e}
\usepackage{rotating}
\usepackage{multirow}
\usepackage{mathtools} 
\usepackage{graphicx}

\usepackage[T1]{fontenc}
\usepackage[utf8]{inputenc}
\usepackage{authblk}

\newcommand*{\bfrac}[2]{\genfrac{\lbrace}{\rbrace}{0pt}{}{#1}{#2}}

\newtheorem{definition}{Definition}

\begin{document}

\title{Next Basket Prediction using Recurring Sequential Patterns}

\author[1,2]{Riccardo Guidotti\thanks{riccardo.guidotti@di.unipi.it}}
\author[1,2]{Giulio Rossetti\thanks{giulio.rossetti@di.unipi.it}}
\author[1,2]{Luca Pappalardo\thanks{luca.pappalardo@di.unipi.it}}
\author[2]{Fosca Giannotti\thanks{fosca.giannotti@isti.cnr.it}}
\author[1]{Dino Pedreschi\thanks{dino.pedreschi@di.unipi.it}}
\affil[1]{Department of Computer Science, University of Pisa, Italy}
\affil[2]{ISTI-CNR, Pisa, Italy}

\maketitle

\begin{abstract}
Nowadays, a hot challenge for supermarket chains is to offer personalized services for their customers. 
\emph{Next basket prediction}, i.e., supplying the customer a shopping list for the next purchase according to her current needs, is one of these services.
Current approaches are not capable to capture at the same time the different factors influencing the customer's decision process: co-occurrency, sequentuality, periodicity and recurrency of the purchased items.
To this aim, we define a pattern \emph{Temporal Annotated Recurring Sequence} (\emph{TARS}) able to capture simultaneously and adaptively all these factors. 
We define the method to extract TARS and develop a predictor for next basket named \emph{TBP} (\emph{TARS Based Predictor}) that, on top of TARS, is able to to understand the level of the customer's stocks and recommend the set of most necessary items.
By adopting the TBP the supermarket chains could crop tailored suggestions for each individual customer which in turn could effectively speed up their shopping sessions.
A deep experimentation shows that TARS are able to explain the customer purchase behavior, and that TBP \mbox{outperforms the state-of-the-art competitors.} 

\end{abstract}

\section{Introduction}
\label{sec:introduction}
%
%
%
%
%
%
%
%
%
%

Detecting the purchase habits of customers and their evolution in time is a crucial challenge 
for effective marketing policies and engagement strategies. 
In such context one of the most promising facilities retail markets can offer to their customers is \emph{next basket prediction}, i.e., the automated forecasting of the next basket that a customer will purchase.
Indeed, an effective basket recommender can act as a \emph{shopping list reminder} for a customer, suggesting the items that she could probably need.

A successful realization of this application requires an in-depth knowledge of an individual's general and recent behavior \cite{mittal1996role}. 
In fact, purchasing patterns of individuals evolve in time and can experience deep changes due to both environmental reasons, like seasonality of products or retail policies, and personal reasons, like diet changes or shift in personal status or preferences. 
As consequence, a satisfactory solution to next basket prediction must be highly \emph{adaptive} to the evolution of a customer's behavior, the recurrence of her purchase patterns and their periodic changes.

In this paper we propose the \emph{Temporal Annotated Recurring Sequences} (\emph{TARS}), adaptive patterns which model the purchasing behavior of an individual by four main characteristics. 
Firstly TARS consider the \emph{co-occurency}: a customer systematically purchases a set of items together.
Secondly TARS model the \emph{sequentiality} of purchases, i.e., the fact that a customer systematically purchases a set of items after another one. 
Third TARS consider \emph{periodicity}: a customer can systematically make a sequential purchase only in specific periods of the year, because of environmental factors or personal reasons. 
Fourth, TARS consider the \emph{recurrency} of a sequential purchase during each period, i.e., how frequently that sequential purchase appears during a customer's period of the year. 
Modeling these four aspects -- co-occurence, sequentiality, periodicity and recurrency -- is fundamental in our opinion to detect the behavior of an individual and its evolution in time. 
On one hand future needs depend on the needs already satisfied: what a customer will purchase depends on what she already purchased last time. 
On the other hand, the needs of a customer depends on her specific habits, i.e., recurring purchases she makes over and over again. 
However habits are far from being static, since they are affected by both endogenous and personal factors \cite{dagher2007shopping,knutson2007neural}. 
Therefore, periodicity is a crucial characteristic of an adaptive model for next basket prediction.
We exploit the TARS and the multiple factors they are able to capture for constructing a parameter-free \emph{TARS Based Predictor} (\emph{TBP}).
TBP is able to solve the next basket prediction problem and to provide a reliable list of items to be reminded in the next purchase as basket recommendation.

We demonstrate the effectiveness of our approach by extracting the TARS for thousands of customers in three real-world datasets, including unique datasets covering a seven years long period.
We show how TARS are easily readable and interpretable, a characteristic which allows to gain useful insights about the purchasing patterns of products and customers. 
Then, we implement a repertoire of state-of-the-art methods and compare them with TBP. 
Our results show that \emph{(i)} TBP outperforms the competitors, 
 \emph{(ii)} it is able to predict up to the next 20 baskets, and \emph{(iii)} the quality of its predictions stabilizes after about 36 weeks.

Finally, it is worth underlining that both TARS and TBP are markedly \emph{user-centric} approaches, in the sense that they use just the data of a customer to make predictions about that customer \cite{pentland2011personal,kalapesi2013unlocking}. 
This aspect eases the customers' personal data management and allows for developing tailored basket recommenders that can even run on the customers' mobile devices \cite{de2014openpds,vescovi2014my}.

In summary, our contributions are the following:
\emph{(i)} we introduce TARS and a parameter-free algorithm to extract them 
(Section \ref{sec:method});
\emph{(ii)} we develop TBP, a predictor based on TARS able to 
produce a shopping list reminder (Section \ref{sec:predictor});
\emph{(iii)} we extract TARS from real-world datasets and show how they are easily interpretable and readable (Section \ref{sec:experiments});
\emph{(iv)} we characterize TBP and compare it with state-of-the-art methods on real datasets (Section \ref{sec:experiments}).
The rest of the paper is organized as follows.
Section \ref{sec:related} reviews existing approaches and Section \ref{sec:problem} formalizes the problem. 
Finally, Section \ref{sec:conclusion} concludes the paper suggesting future research directions.

\section{Related Work}
\label{sec:related}

Next basket prediction is mainly aimed at the construction of effective recommender systems (or recommenders). 
Recommenders can be categorized into \emph{general}, 
\emph{sequential}, 
\emph{pattern-based} 
and \emph{hybrid}. 
General recommenders are based on collaborative filtering and produce recommendations 
with respect to general customers' preferences \cite{su2009survey}. 
Sequential recommenders are based on Markov chains and produce recommendations 
exploiting sequential information and recent purchases \cite{chand2012sequential}. 
Pattern-based recommenders base predictions on frequent itemsets extracted from the purchase history of all customers \mbox{while discarding sequential information \cite{hsu2004mining,lazcorreta2008towards}.} 


The hybrid approaches combine the ideas underlying general and sequential recommenders.
In \cite{rendle2010factorizing} the authors use personalized transition graphs over Markov chains and compute the probability that a customer will purchase an item by using an optimization criteria named Bayesian Personalized Ranking \cite{rendle2009bpr}. 
HRM \cite{wang2015learning} and DREAM \cite{yu2016dynamic} exploit both general customers' preferences and sequential information by using recurrent neural networks.  
A different hybrid approach is described in \cite{wang2014modeling} where is developed a probability model by merging Markov chain and association patterns.

All the approaches described above suffer from several limitations. 
General recommenders and pattern-based recommenders do not take into account neither the sequential information (i.e., which item is bought after which) nor the customers' recency. 
On the other hand, sequential recommenders assume the independence of items in the same basket and do not capture factors like mutual influence.  
Furthermore, all of them require transactional data about many customers in order to make a prediction for a single customer.
For this reason, they do not follow the \emph{user-centric} vision for data protection as promoted by the World Economic Forum \cite{kalapesi2013unlocking,pentland2011personal}, which incentives personal data management for every single user of a data-based service.
Cumby et al. \cite{cumby2004predicting} propose a basket predictor which embraces the user-centric vision by reformulating next basket prediction as a classification problem: they build a distinct 
classifier for every customer and for every item hence performing predictions by relying just on her personal data. 
However, this approach also assumes the \mbox{independence of items purchased together.}


Finally, the main drawback of the existing hybrid approaches \cite{yu2016dynamic,wang2015learning,wang2014modeling}
is that their 
predictive models 
are hardly readable and interpretable by humans. 
Interpretability of a predictive model, i.e., the possibility to understand the mechanisms underlying the predictions \cite{ribeiro2016why}, is highly valuable for a retail chain manager 
interested in interpreting the predictive model to improve the marketing strategies and the service offered. 
Moreover, interpretability is also important to the customers who want to gain insights about \mbox{their personal purchasing behavior.} 


%
%


\section{Next Basket Prediction Problem}
\label{sec:problem}
We refer to \emph{next basket prediction} as the task of predicting which items a customer will purchase in her next transaction. 
Formally,  let $C = \{ c_1, \dots, c_z\}$ be a set of $z$ customers and $I = \{ i_1, \dots, i_m\}$ be a set of $m$ items. 
Given a customer $c$, $B_c = \langle b_{t_1}, b_{t_2}, \dots, b_{t_n} \rangle$ is the ordered \emph{purchase history} of her baskets (or transactions), where $b_{t_i} \subseteq I$ represents the basket composition and $t_i \in [t_1, t_n]$ is the transaction time. 
We indicate with $\mathcal{B} = \{B_{c_1}, B_{c_2}, \dots, B_{c_{z}} \}$ is the set of all customers' purchase histories.

Given the purchase history $B_c$ of customer $c$ and the time $t_{n+1}$ of the next transaction, next basket prediction consists in providing the set $b^*$ of $k$ items that 
$c$ will \mbox{purchase in the next transaction $b_{t_{n+1}}$.}

\vspace{1mm}
 
Our approach to next basket prediction aims at overcoming the main limitations of existing methods illustrated in Section \ref{sec:related}.
To this purpose, we propose a hybrid predictor which combines ideas underlying sequential and pattern-based recommenders. 
The approach consists of two main components.
The first one is the extraction of \emph{Temporal Annotated Recurring Sequences} (\emph{TARS}) from the customer's purchase history, i.e., sequential recurring patterns able to capture the customer's purchasing habits.
The second one is in the \emph{TARS Based Predictor} (\emph{TBP}), a predictive method that exploits the TARS of a customer to forecast her next basket.


\section{Capturing Purchasing Habits}
\label{sec:method}

In this section we formalize TARS and describe how to extract them form the purchase history of a customer.

\subsection{Temporal Annotated Recurring Sequences}
\label{sec:model}
\emph{Temporal Annotated Recurring Sequences} (\emph{TARS}) model recurrent sequential purchases of a customer, i.e., the fact that a set of items are typically purchased together, the fact that that a set of items is typically purchase after another set of items, and the recurrence of the sequential purchase, i.e., when and how often it occurs in the purchase history of the customer. 
We define a TARS as follows:

\begin{definition}[Temporal Annotated Recurring Sequence]
Given the purchase history $B$ of a customer, a \emph{temporally annotated recurring sequence} (\emph{TARS}) is a quadruple $\gamma {=} (S, \alpha, p, q)$, where $S {=} \langle X, Y \rangle$ ${=} X \rightarrow Y$ is the \emph{sequence} of itemsets, $\alpha {=} (\alpha_1, \alpha_2) \in \mathbb{R}_+^2$, $\alpha_1 \le \alpha_2$ is the \emph{temporal annotation}, $p$ is the \emph{number of periods} in which the sequence recurs, and $q$ is the \emph{median of the number of occurrences in each period}. 
A TARS will also be represented as follows:
$$\gamma = X \xrightarrow[p, q]{\alpha} Y$$
\end{definition}

We refer to $\Gamma_c = \{ \gamma_1, \dots, \gamma_m \}$ as the set of all the TARS of a customer $c$. 
A TARS is based on the concept of \emph{sequence}, ${S = \langle X, Y \rangle =}$ $X \rightarrow Y$, which intuitively indicates that the an itemset $Y$ is typically purchased after another itemset $X$.
The itemsets themselves point out which items are purchased together.
For example, a sequence $\{a\} \rightarrow \{b, c\}$ indicates that the items $\{ b,c \}$ are purchased together after the itemset $\{a\}$. 
The temporal annotation $\alpha = (\alpha_1, \alpha_2)$ indicates the minimum intra-time $\alpha_1$ and maximum intra-time $\alpha_2$ \emph{intra-time} of the sequence, i.e., the range of time elapsing between the purchase of $X$ and the purchase of $Y$. 
A sequence can appear in several distinct \emph{periods}, i.e., time intervals where the sequence occurs continuously. 
The number of periods $p$ characterizes these recurrences, that is, in how many periods the sequence $S$ appears. 
Finally, $q$ indicates how many times $S$ typically occurs in a period.

\begin{definition}[Sequence]
Given the purchase history of a customer $B_c = \langle b_{t_1},$ $\dots, b_{t_n} \rangle$, we call $S = \langle X, Y \rangle = X \rightarrow Y$ a \emph{sequence} if the pair of itemsets $X \subseteq b_{t_h}$ and $Y \subseteq b_{t_l}$, $X, Y \neq \emptyset$, $t_h < t_l$ and ${\nexists \; S' = X' \rightarrow Y'}$, ${X' \subseteq b_{t'_h}}$ and ${Y' \subseteq b_{t'_l}}$ such that $t'_h, t'_l \in [t_h, t_l]$. 
$X$ and $Y$ are called the \emph{head} and the \emph{tail} of the sequence, respectively. 
\end{definition}

We denote with $T_S = \langle t_{j_1}, \dots, t_{j_m} \rangle$ the \emph{head time list} of $S$, i.e., the ordered list of the head's time of all the occurrences of $S$ in the purchase history of the customer.
The \emph{support} $|T_S|$ of a sequence $S$ is the size of its head time list.
We call \emph{length of a sequence} $|S| = |X| + |Y|$ the sum of sizes of the head and of the tail. 
We say that a sequence $S'$ is a \emph{subsequence} of $S''$, $S' \; \vec{\subseteq} \; S''$ if ${X' \subseteq X'' \wedge Y' \subseteq Y''}$.

\begin{definition}[Intra-Time]
We define $\alpha_h {=} t_l {-} t_h$ as the \emph{intra-time} of an occurrence of a sequence $S$, i.e., the difference between the time of the head and the time of the tail.
We denote with $A_{S} {=} \langle \alpha_1,$ $\dots, \alpha_m \rangle$ the ordered intra-time list of all the \mbox{occurrences of $S$ in $B$.}
\end{definition} 

\begin{definition}[Inter-Time]
Given the head time list $T_S$, we define $\delta_j = t_{l_i} - t_{l_j}$ with ${t_{l_i}, t_{l_j} \in T_{S}}$ and ${t_{l_i} < t_{l_j}}$ as the \emph{inter-time} of a sequence $S$, i.e., the difference between the times of the heads of two consecutive occurrences of $S$. 
We denote with $\Delta_{S} {=} \langle \delta_{1}, \dots, \delta_{m} \rangle$ the ordered inter-time list of $S$. 
\mbox{We impose $\delta_{m} {=} \alpha_{m}$ by construction.}
\end{definition}

\begin{figure}[tb]
    \centering
   	\includegraphics[trim = 0mm 0mm 0mm 3mm, clip, width=0.75\linewidth]{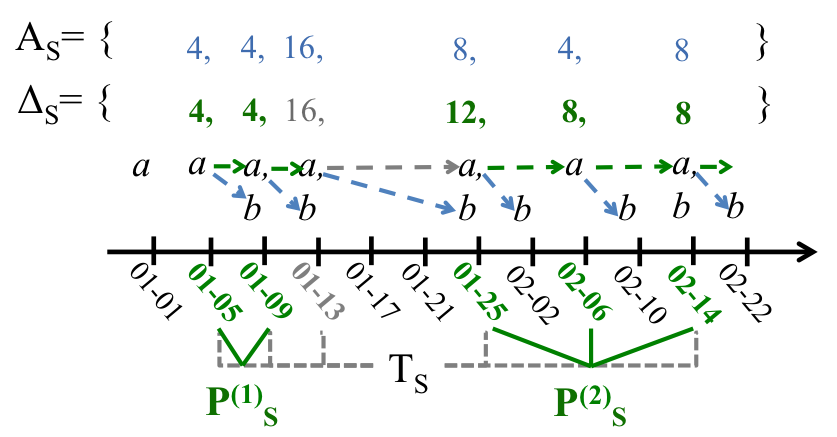}
	\vspace{-4mm}
	\caption{Head time list $T_S$, intra-time list $A_S$, inter-time list $\Delta_S$ and periods $P_S^{(1)}$, $P_S^{(2)}$ of sequence $S {=} \{a\} {\rightarrow} \{b\}$.}
\label{fig:toysample1}
\end{figure}

To clarify the concepts defined above, let us consider the example in Table~\ref{tab:toysample} which shows the purchase history of a customer. 
Based on the example, Figure~\ref{fig:toysample1} shows the occurrences of sequence $\mathit{S = \{a\} \rightarrow \{b\}}$. 
The head time list $T_S$ consists of the times of the heads of all the occurrences of $S$, hence $T_{S} = \langle 01\mbox{-}05, 01\mbox{-}09, 01\mbox{-}13,$ $01\mbox{-}25, 02\mbox{-}06, 02\mbox{-}14 \rangle$. 
The intra-time list $A_S$ consists of the differences between the heads and the tails of all the occurrences of $S$, hence ${A_{S} = \langle 4, 4, 16, 8, 4, 8\rangle}$. 
The inter-time list $\Delta_S$ consists of all differences between the head times of two consecutive sequences, hence ${\Delta_{S} = \langle 4, 4, 16, 12, 8, 8 \rangle}$. 
Note that: \emph{(i)} for each $t_j \in T_{S}$ we have that $\alpha_j \leq \delta_j$, i.e., the intra-time of a sequence is always lower or equal than its inter-time; \emph{(ii)} for $S = X \rightarrow X$, we have $A_{S} = \Delta_{S}$.

\begin{table}[tb]
\centering
\begin{tabular}{|c|c|c|c|c|}
\cline{0-1}\cline{4-5}
timestamp  & basket        &  & timestamp  & basket                 \\
\cline{0-1}\cline{4-5}
01-01 & $a, b, g, h$    &  & 01-25 & $a, b, c, g, h$          \\
01-05 & $a, c, d$       &  & 02-02 & $b, c, d$                   \\
01-09 & $a, b, e, f, h$ &  & 02-06 & $a, c, d, e, f, i$          \\
01-13 & $a, b, c, d, h$ &  & 02-10 & $b, e, f, h$          \\
01-17 & $c, d, e, f, g$ &  & 02-14 & $a, b, c, d, e, f, g, h$ \\
01-21 & $e, f, g$       &  & 02-22 & $a, b, g, h, i$         \\
\cline{0-1}\cline{4-5}
\end{tabular}
\caption{Example of customer purchase history $B_c$.}
\label{tab:toysample}
\vspace{-6mm}
\end{table}

\begin{definition}[Period] 
Given a maximum inter-time $\delta^{max}$, a minimum number of occurrences $q^{min}$, the head time list $T_S$ and the inter-time list $\Delta_S$ of a sequence $S$, we call \emph{period} an ordered time list $P_S^{(j)} = \langle t_{h}, \dots, t_{l} \rangle \subseteq T_{S}$ such that $\forall \; t_w \in P_S^{(j)}$, $\delta_w \leq \delta^{max}$, $P_S^{(j)}$ is maximal, i.e., $\delta_{h-1} > \delta^{max}$, $\delta_{l+1} > \delta^{max}$, and $|P_S^{(j)}| \geq q^{min}$.
We denote with $P_{S} = \{ P_S^{(1)}, \dots, P_S^{(m)} \}$ the set of periods of $S$.
\end{definition}

The period of a sequence $S$ captures a temporal interval in which $S$ occurs at least $q^{min}$ times and the time between any two occurrences is at most $\delta^{max}$. 
The support $|P_S^{(j)}|$ of a period indicates how many times $S$ occurs in $P_S^{(j)}$.
In the example of Figure~\ref{fig:toysample1}, for $\delta^{max} = 14$ and $q^{min} = 2$ we have two periods $P_S^{(1)} = \langle 01\mbox{-}05, 01\mbox{-}09 \rangle$ and $P_S^{(2)} = \langle 01\mbox{-}25, 02\mbox{-}06, 02\mbox{-}14 \rangle$ with support 2 and 3 respectively.

\begin{definition}[Recurring Sequence]
Let $P_{S} = \{ P_S^{(1)}, \dots, P_S^{(m)} \}$ be a set of periods, we define $rec(S) {=} |P_{S}|$ as the \emph{recurrence} of $S$, i.e., the number of periods $P_S$ in the purchase history. 
Given a minimum number of periods $p^{min}$, {$S$ is a \emph{recurring sequence} if $rec(S) \geq p^{min}$.}
\end{definition}

In the example of Figure~\ref{fig:toysample1}, for $p^{min} {=} 2$ we have $rec(S) {=} 2$ meaning that $S$ is recurring. 
By specifying the maximum inter-time $\delta^{max}$, the minimum number of occurrences $q^{min}$, and the minimum number of periods $p^{min}$, we can determine the set $\Gamma_c$ of TARS that can be extracted from the purchase history $B_c$ a customer $c$.

TARS are an evolution of both recurring patterns \cite{kiran2015discovering} and temporally annotated sequences \cite{giannotti2006efficient}:
the former models recurrency but do not model sequentiality and periodicity, while the latter models sequentiality and periodicity but do not model recurrency.
TARS, besides co-occurrence, fills the gaps \mbox{by modeling all the three aspects.}

{

\setlength{\textfloatsep}{2.5mm}
\setlength{\intextsep}{2.5mm}
\setlength{\floatsep}{2.5mm}

\subsection{TARS Extraction Procedure}
\label{sec:tars_algorithm}
To extract the TARS from a customer's purchase history we use an extension of the well-known \emph{FP-Growth} algorithm \cite{han2000mining}. 
It builds a \emph{FP-tree} which captures the frequency at which itemsets occur in the dataset. 
It has been shown in literature \cite{amphawan2010mining,fournier2016phm,kiran2015finding} that FP-Growth can be extended by attaching additional information to a FP-tree node in order to calculate the desired type of pattern. 

In our approach we extend the FP-tree into a \emph{TARS-tree}. 
Every node of a TARS-tree stores a sequence $S$, the time list $T_{S}$, its support $|T_{S}|$, the intra-time list $A_S$, the inter-time list $\Delta_S$ and the periods $P_{S}$ derived from $T_S$ with respect to $\delta^{max}$ and $q^{min}$.

The TARS extraction procedure is described in Algorithm~\ref{alg:extract_tars}.
In the first step it extracts from the purchase history $B$ the \emph{base sequences} $\mathcal{S}$, i.e., the sequences of length 2 (line 1).
Then, a set of parameters $\{\delta^{max}_S\}, \{q^{min}_S\}, \{p^{min}_S\}$ is estimated for each base sequence $S \in \mathcal{S}$ with respect to $B$ (line 2).
The base sequences $\mathcal{S}$ are then filtered with respect to these parameters and the base recurring sequences $\mathcal{S}^*$ are extracted, while the other base sequences are discarded to reduce the search space (line 3).
Finally, the TARS-tree $\Psi$ is built on the base recurring sequences $\mathcal{S}^*$ (line 4), and the set $\Gamma$ of TARS annotated with $\alpha, p, q$ is extracted from the TARS-tree $\Psi$ (line 5) according to the FP-Growth procedure.

\begin{algorithm2e}[tb]
\caption{$extractTars(B)$}
\label{alg:extract_tars}
$\mathcal{S} \leftarrow extractBaseSequences(B)$\;
$\{\delta^{max}_S\}, \{q^{min}_S\}, \{p^{min}_S\} \leftarrow parametersEstimation(B, \mathcal{S})$\;
$\mathcal{S}^* \leftarrow sequenceFiltering(B, \mathcal{S}, \{\delta^{max}_S\}, \{q^{min}_S\}, \{p^{min}_S\})$\;
$\Psi \leftarrow buildTars\mbox{-}Tree(B, \mathcal{S}^*, \{\delta^{max}_S\}, \{q^{min}_S\}, \{p^{min}_S\})$\;
$\Gamma \leftarrow extractTarsFromTree(\Psi)$\;
\Return{$\Gamma$}\;
\end{algorithm2e}

\subsubsection{Data-Driven Parameters Estimation}
\label{sec:parameters}
In order to make the parameters $\delta^{max}$, $q^{min}$, $p^{min}$ adaptive not only to the individual customer \cite{keogh2004towards}, but also to the various sequences in $B_c$, 
we apply two pre-processing steps on the base sequences (lines 1--2 Algorithm~\ref{alg:extract_tars}).
The first pre-processing step is the data-driven estimation of the sets of parameters $\{ \delta^{max}_S \}$, $\{ q^{min}_S \}$, $\{ p^{min}_S \}$ \mbox{described in Algorithm~\ref{alg:paramest}.}

Let $\mathcal{S}$ be the set of base sequences and $\hat{\delta_{S}}$ be the median of inter-times in $\Delta_{S}$ (Algorithm~\ref{alg:paramest}, line 2). 
Given a base sequence $S$, we estimate parameter $\delta^{max}$ by the following two steps: 
\emph{(i)} we group the base sequences with similar inter-times $\hat{\delta_{S}}$ (line 3) obtaining a set of clusters $\mathcal{C}_{\delta^{max}} {=} \{C_1, \dots, C_v\}$;
\emph{(ii)} if $S \in C_h$, $C_h \in \mathcal{C}_{\delta^{max}}$, we set $\delta^{max}_S$ as the median of the $\hat{\delta_{S}}$ values 
in cluster $C_h$ (lines 4--5). 

Then, we calculate the periods $TC_S$ compliant only with the temporal constraint $\delta^{max}_S$ (lines 6--8) and we estimate 
$\{ q^{min}_S \}$ as follows: 
\emph{(i)} we group the base sequences with similar median number of occurrences per period $\hat{q_{S}}$, 
producing a set of clusters $\mathcal{C}_{q^{min}} = \{C_1, \dots, C_g \}$ (line 9); 
and \emph{(ii)} if $S \in C_h$, $C_h \in \mathcal{C}_{q^{min}}$ we set $q^{min}_S$ as the median of the $\hat{q_{S}}$ in $C_h$ (lines 10--11).

Similarly, we estimate $\{ p^{min}_S \}$ as follows: 
\emph{(i)} we compute the sum of the number of occurrences of a base sequence in the periods $w_S$ and we calculate the expected number of occurrences per period $e_S$ as $w_S / |P_{S}|$ (lines 12--14); 
\emph{(ii)} we group the base sequences with similar $e_S$ producing a set of clusters $\mathcal{C}_{p^{min}} = \{ C_1, \dots, C_d \}$ (line 15); 
and \emph{(iii)} if $S \in C_h$, $C_h \in \mathcal{C}_{p^{min}}$, we set $p^{min}_S$ as the median of the number of periods of the base sequences in $C_h$ (lines 16--17).

We group the base sequences by dividing the values 
into equal-sized bins \cite{pearson1894contributions}, whose number is estimated as the maximum between the estimated number of bins suggested by the Sturges \cite{sturges1926choice} and the Freedman-Diaconis methods \cite{freedman1981histogram}.

\begin{algorithm2e}[tb]
\caption{$parametersEstimation(\mathcal{S}, B)$}
\label{alg:paramest}
$D_{\delta^{max}} \leftarrow \emptyset$; $D_{q^{min}} \leftarrow \emptyset$; $D_{p^{min}} \leftarrow \emptyset$\;
\lForEach{$S \in \mathcal{S}$}{$D_{\delta^{max}} \leftarrow D_{\delta^{max}} \cup \{ \hat{\delta_{S}} = median(\Delta_{S}) \}$}
$\mathcal{C}_{\delta^{max}} \leftarrow groupSimilar(D_{\delta^{max}})$\;
\For{$C_h \in \mathcal{C}_{\delta^{max}}$}{
	\lForEach{$S \mbox{ } assignedTo(C_h)$}{$\delta^{max}_S \leftarrow median(C_h)$}
}
\For{$S \in \mathcal{S}$}{
	$\mathit{TC}_S \leftarrow getTemporalyCompliantPeriods(S, B, \{ \delta^{max}_S \})$\;
	$D_{q^{min}} \leftarrow D_{q^{min}} \cup \{ median(\{ \hat{q_{S}} = |\mathit{TC}^{(j)}_S| \: \mbox{s.t.} \: \mathit{TC}^{(j)}_S {\in} \mathit{TC}_S \}) \}$\;
}
$\mathcal{C}_{q^{min}} \leftarrow groupSimilar(D_{q^{min}})$\;
\For{$C_h \in \mathcal{C}_{q^{min}}$}{
	\lForEach{$S \mbox{ } assignedTo(C_h)$}{$q^{min}_S \leftarrow median(C_h)$}
}
\For{$S \in \mathcal{S}$}{
	$P_S \leftarrow getPeriods(S, B, \{\delta^{max}_S \}, \{q^{min}_S \})$\;
	$w_S \leftarrow \sum_{P^{(j)}_S \in P_S}|P^{(j)}_S|$;
	$e_S \leftarrow w_S / |P_S|$;
	$D_{p^{min}} \leftarrow D_{p^{min}} \cup \{ e_S \}$\,
}
$\mathcal{C}_{p^{min}} \leftarrow groupSimilar(D_{p^{min}})$\;
\For{$C_h \in \mathcal{C}_{p^{min}}$}{
	\For{$S \mbox{ } assignedTo(C_h)$}{
		${p^{min}_S \leftarrow median(\{ rec(P_{S'}) {=} |P_{S'}| \mbox{s.t.} S' assignedTo(C_h) \})}$\;
	}
}

\Return{$\{ \delta^{max}_S \}, \{ q^{min}_S \}, \{ p^{min}_S \}$}\;
\end{algorithm2e}

\subsubsection{Sequence Filtering}
The second pre-processing step consists in the selection of the \emph{base recurring sequences}, i.e., the base sequences satisfying the sets of parameters $\{ \delta^{max}_S \}$, $\{ q^{min}_S \}$, $\{ p^{min}_S \}$.
We apply this filtering in order to reduce the search space so that the building of the TARS-tree and the TARS extraction (lines 4--5 Algorithm~\ref{alg:extract_tars}) are employed only on the super-sequences of the base recurring sequences.
In other words, if $S_1$ is not a base recurring sequence and $S_1 \; \vec{\subseteq} \; S_2$, then we assume as heuristic that $S_2$ is not recurring too, and we eliminate it through the sequence filtering process.
We adopt the sequence filtering heuristic for reducing the search space because the \emph{antimonotonic property} \cite{agrawal1993mining} does not apply to TARS.
Consider $S_1 {=} \{c\} {\rightarrow} \{c\}$ and $S_2 {=} \{c, d\} {\rightarrow} \{c\}$ in the example of Table~\ref{tab:toysample}. 
We have that $S_1 \; \vec{\subseteq} \; S_2$.
Given $\delta^{max} {=} 14$, $q^{min} {=} 2$ and $p^{min} {=} 2$, we have $rec(S_1) {=} 1$ and $rec(S_2) {=} 2$. 
Hence, $S_2$ is recurrent while $S_1$ is not, and the anti-monotonic property is not satisfied.

However, it is clear from this example that a TARS with $S_1$ could be useful for the prediction because, despite $rec(S_1) = 1$ in total it occurs six times $|P^{(1)}_{S_1}| = 6$. 
In real-world, $\{ c \}$ could be a fresh product (like milk or salad) that is repeatedly and frequently purchased. 
Hence, an imposed parameter setting could be not appropriate because \emph{(i)} it could remove too many TARS which are in fact useful for the prediction; \emph{(ii)} it could consider too many valid base sequences and not prune enough the search space.

\vspace{1mm}
For these reasons we developed the two pre-processing steps heuristic for parameters estimation described in this section.

\begin{algorithm2e}[tb]
\caption{$getActiveTARS(B, t_{n+1}, \Gamma)$}
\label{alg:active_tars}
$\hat{\Gamma} \leftarrow \emptyset$;
$Q \leftarrow \emptyset$;
$L \leftarrow \emptyset$;
$\Upsilon \leftarrow \Gamma$\;
\For{$b_{t_j}, b_{t_{j-1}} \in \mbox{sort-desc}(B)$}{
	$\alpha_{j-1} \leftarrow t_j - t_{j-1}$\;
	\For{$X \subseteq b_{t_{j-1}}$}{
		\For{$Y \subseteq b_{t_j}$} {
			\If{$\exists \; \gamma \in \Upsilon \; | \; {\gamma = (S, \alpha, p, q)} \wedge {\alpha_1 \leq \alpha_{j-1} \leq \alpha_2} \wedge {\; \; \; \: S = \langle X, Y \rangle = X \rightarrow Y}$}{
				\uIf{$\gamma \in \hat{\Gamma}$}{
					$Q_{\gamma} \leftarrow Q_{\gamma} + 1$;
					$L_{\gamma} \leftarrow t^{j-1}$\;
					\lIf{$Q_{\gamma} > q$}{
						$\hat{\Gamma} \leftarrow \hat{\Gamma} / \{ \gamma \}$;
						$\Upsilon \leftarrow \Upsilon / \{ \gamma \}$}
					\lIf{$L_{\gamma} - t_{j-1} {>} q \cdot (\alpha_1 \mbox{-} \alpha_2)$}{
						$\Upsilon \leftarrow \Upsilon / \{ \gamma \}$}
				}\Else{
					$\hat{\Gamma} \leftarrow \hat{\Gamma} \cup \{ \gamma \}$;
					$Q_{\gamma} \leftarrow 1$;
					$L_{\gamma} \leftarrow t_{j-1}$\;
				}
				\lIf{$\Upsilon = \emptyset$}{\Return{$\hat{\Gamma}, Q$}}
			}
		}
	}
}
\Return{$\hat{\Gamma}, Q$}\;
\end{algorithm2e}

\section{TARS Based Predictor}
\label{sec:predictor}
On top of the set $\Gamma_c$ of TARS extracted from the purchase history $B_c$ of customer $c$ we build the \emph{TARS Based Predictor} (\emph{TBP}), an approach for next basket prediction that is markedly \emph{personalized} and \emph{user-centric} \cite{kalapesi2013unlocking,pentland2011personal}, in the sense that just the model build on the individual purchase history $B_c$ of customer $c$, i.e., her TARS $\Gamma_c$, is used to make the predictions about that customer $c$.

TBP exploits TARS to simultaneously embed complex item interactions such as the co-occurrence (which item is bought with which), sequential relationship (which items are bought after which), periodicity (which item is bought when) and typical times of re-purchase (after when re-purchases happen).
These factors enable TBP to observe the recent purchase history of a customer and understand which are the \emph{active} patterns, i.e., the patterns that the customer is currently following in her purchasing. 
In turn, by realizing which are the active patterns TBP can provide the set of items that she will need at the time of the next purchase.
It is worth noting that TBP is parameter-free: all the parameters of the TARS model $\Gamma_c$ are automatically estimated for each customer on her personal data $B_c$, avoiding the usual case where the same parameter setting is used indiscriminately for all the customers \cite{keogh2004towards}.

\vspace{1mm}
Given the purchasing history $B_c$ of customer $c$, the time $t_{n+1}$ of $c$'s next transaction, and $c$'s TARS set $\Gamma_c$, the TBP approach works in two steps. 
First, it selects the set $\hat{\Gamma}_c$ of \emph{active} TARS. 
Second, it computes a score $\Omega_{c_i}$ for every item $i$ belonging to a TARS in $\hat{\Gamma}_c$, ranks the items according to their score $\Omega_{c_i}$, and selects the top $k$ items as the basket prediction for customer $c$.

\vspace{1mm}
Algorithm~\ref{alg:active_tars} shows the procedure of the TBP to select the \emph{active} TARS of a customer $\hat{\Gamma}$. 
First, it sorts the purchase history $B$ ordering it chronologically from the most recent basket to the oldest one, then it loops on pairs of consecutive baskets (line 2) searching for a set $\Upsilon$ of \emph{potentially active} TARS (lines 4--7).
When it finds a potentially active TARS $\gamma$, it considers two cases.
If the sequence $S$ of $\gamma$ is encountered for the first time, the algorithm adds $\gamma$ to the set $\hat{\Gamma}$ of active TARS and initializes two variables: the number of times $\gamma$ has been encountered $Q_{\gamma}$ and its last starting time $L_\gamma$ (line 13). 
In the second case, the algorithm increments $Q_\gamma$ and updates $L_\gamma$ (line 9). 
If $Q_\gamma > q$ the algorithm removes $\gamma$ from the set of active TARS and from the set of potentially active TARS (line 9).
If too much time has passed between the last beginning of TARS $\gamma$ and its next occurrence (line 11), the algorithm does not look for that TARS $\gamma$ anymore and removes it from $\Upsilon$.
Algorithm~\ref{alg:active_tars} stops either when the set of potentially active TARS is empty (line 14), or when the entire purchase history $B$ has been scanned (line 15).
Finally, it returns the set $\hat{\Gamma}$ of active TARS and the number of times $Q$ the sequences of the active TARS have occurred in the last period.

\begin{algorithm2e}[tb]
\caption{$calculateItemScore(B, \hat{\Gamma}, Q)$}
\label{alg:calculate_score}
$\Omega \leftarrow \emptyset$;
\lForEach{$i \in I$}{$\Omega_i \leftarrow 0$}
\For{$\gamma = (S = \langle X, Y \rangle, \alpha, p, q) \in \hat{\Gamma}$}{
	\lForEach{$i \in Y$}{$\Omega_i \leftarrow \Omega_i + (q - Q_{\gamma})$}
}
\For{$i \in \{ i \; | \; \exists \; \gamma = (S = \langle X, Y \rangle, \alpha, p, q) \in \hat{\Gamma}, i \in Y \}$}{
	$\Omega_i \leftarrow \Omega_i + sup(i)$
}
\Return{$\Omega$}\;
\end{algorithm2e}

\vspace{1mm}
Algorithm~\ref{alg:calculate_score} shows the procedure of TBP to compute the items' scores.
First, it sets to zero the score of each item $\Omega_i$ (line 1)
Then, for every active TARS $\gamma$ containing item $i {\in} Y$, it increases $\Omega_i$ with the difference between the typical number of occurrences $q$ of $\gamma$ and $Q_\gamma$ indicating the number of times that the sequence of $\gamma$ occurred in the recent 
history (lines 2--3). 
Finally, $\Omega_i$ is augmented with the support of item $i$ for the items in \mbox{the tail of the active TARS (lines 4--5).}

After this procedure TBP ranks the items' scores $\Omega_c$ in descending order and returns the top-$k$ items as the predicted basket. 

}

\section{Experiments on Retail Data}
\label{sec:experiments}
In this section, we report the experiments performed on three real-world datasets in order to show the properties of the TARS and the effectiveness of TBP in next basket prediction.

State-of-the-art methods \cite{rendle2010factorizing,wang2015learning,yu2016dynamic,cumby2004predicting} fix the size of the predicted basket to $k {=} 5$ or $k {=} 10$. 
However, we think that the size $k$ of the predicted basket should adapt to the customer's personal behavior. 
Indeed, if a customer typically purchases baskets with a few items it is useless to predict a basket with a large number of items. 
On the other hand, if a customer typically purchases baskets with a large number of items, the prediction of a small basket will not cover most of the items purchased. 
In this paper we report the evaluation of the predictions made using both a fixed length $k \in [2, 20]$ for all the customers and using a customer-specific size $k=k_c^*$, where $k_c^*$ indicates the average basket length of customer $c$.

According to the literature, we adopt a \emph{leave-one-out} strategy for model validation \cite{yu2016dynamic,wang2015learning,rendle2010factorizing,cumby2004predicting}: for each customer $c$ we use the baskets in the purchase history $B_c = \{ b_{t_{1}}, \dots, b_{t_{n}}\}$ for extracting the TARS, and the basket $b_{t_{n+1}}$ as test to estimate the performance.

For every customer, we evaluate the agreement of
the predicted basket $b^*$ and the real basket $b$ by using the following 
metrics:

\begin{itemize}
\item \emph{F1-score}, the harmonic \mbox{mean of precision and recall \cite{tan2006introduction}}: 
$$F1\mbox{-}score(b, b^*) = \frac{2 \cdot Precision(b, b^*) \cdot Recall(b, b^*)}{Precision(b, b^*) + Recall(b, b^*)}$$
$$\; \; \; \; \; \; \; \; \; \; \: Precision(b, b^*) = |b \cap b^*|/|b^*| \; \;  Recall(b, b^*) = |b \cap b^*|/|b|$$
\item \emph{Hit-Ratio}, the ratio of customers who received at least one correct prediction (a \emph{hit}) \cite{karypis2001evaluation}:
$$Hit\mbox{-}Ratio(b, b^*) = 1 \mbox{ if } b \cap b^* \neq \emptyset, 0 \mbox{ otherwise}.$$
\item \emph{normalized F1-score}: the F1-score calculated only for the customers having at least one hit.
\end{itemize}

Furthermore, for each customer we compute both \emph{learning time} and \emph{prediction time}. 
The learning time is the amount of time required to extract the model from the data. 
The prediction time is the amount of time the predictor needs to predict the next basket of a customer. 
We perform the experiments on Ubuntu 16.04.1 LTS 64 bit, 32 GB RAM, 3.30GHz Intel Core i7. 




\begin{table}[tb]
\setlength{\tabcolsep}{0.01mm}
\centering
\begin{tabular}{|c|ccccc|}
\hline
\textbf{Dataset} & \multicolumn{1}{c}{\emph{\textbf{cust.}}} & \multicolumn{1}{c}{\emph{\textbf{\# baskets}}} & \multicolumn{1}{c}{\emph{\textbf{\# items}}} & \multicolumn{1}{c}{\emph{\textbf{\begin{tabular}[c]{@{}c@{}}avg basket\\ per cust.\end{tabular}}}}  & \multicolumn{1}{c|}{\emph{\textbf{\begin{tabular}[c]{@{}c@{}}avg basket\\ length\end{tabular}}}} \\
\hline
\textit{Coop-A} & 10,000 & 7,407,056 & 4,594 & $432.4{\pm}353.4$ & $9.4{\pm}5.8$ \\
\textit{Coop-C} & 10,000 & 7,407,056 & 407 & $432.4{\pm}353.4$ & $8.6{\pm}4.9$ \\
\textit{Ta-Feng} & 2,319 & 24,304 & 5,117 & $10.4{\pm}7.5$ & $1.8{\pm}1.1$ \\
\hline
\end{tabular}
\vspace{2mm}
\caption{Statistics of the datasets used in the experiments.}
\label{tab:dataset_stats}
\vspace{-6mm}
\end{table}

\subsection{Datasets}
\label{sec:dataset}
For our experiments we use three real-world transactional datasets: \emph{Coop-A}, \emph{Coop-C} (both extracted from the \emph{Coop} repository) and \emph{Ta-Feng}. 
Table~\ref{tab:dataset_stats} shows the details of the datasets. 

The \emph{Coop} repository is provided by \emph{Unicoop Tirreno}\footnote{https://www.unicooptirreno.it/}, a big retail supermarket chain in Italy. 
It stores 7,407,056 transactions made by 10,000 customers in 23 different shops in the province of Leghorn, over the years 2007-2014. 
The set of Coop items includes food, household, wellness, and multimedia items. 
There are 7,690 different articles classified into 520 market categories. 
From the repository we extract two datasets: \emph{Coop-A} and \emph{Coop-C}. 
The two datasets differ in the items categorization. 
In \textbf{\emph{Coop-A}} (articles) the items of a basket are labeled with a fine-grained categorization which distinguishes, for example, between blood orange and navel orange. 
In \textbf{\emph{Coop-C}} (categories) the items are mapped to a more general category: in the example above blood orange and navel orange are considered the same generic item (orange). 
All the customers in \emph{Coop-A} and \emph{Coop-C} have at least one purchase per month. 

\emph{\textbf{Ta-Feng}}\footnote{http://www.bigdatalab.ac.cn/benchmark/bm/dd?data=Ta-Feng} 
dataset 
covers food, office supplies and furniture, with a total of 23,812 
items. 
It 
contains 817,741 transactions made by 32,266 customers over four months. 
We remove the customers with less than 10 baskets 
and consider \mbox{
the remaining 7\% customers.}

Since we do experiments on real retail datasets we adopt the \emph{day} as time unit, i.e., \mbox{parameters and annotations are expressed in days.}

\subsection{Interpretability of TARS}
The interpretability of TARS is one of the main characteristics of our approach. 
Table~\ref{tab:tars_coop2} shows some examples of TARS extracted from \textit{Coop-C}. 
In the table we report the median of $\alpha$, $p$ and $q$ across all the customers having the presented TARS. 
We observe that TARS with a recurring base sequence are the most supported among the customers. 
For example $\{ milk \} {\rightarrow} \{ milk \}$ and $\{ banana \} {\rightarrow} \{ banana \}$ are supported by more than 90\% of the customers in \emph{Coop-C}. 
The two TARS have similar $q$ (6.58 and 7.20 respectively) indicating that they have similar recurrence degrees, i.e., they occurs a similar number of times in the respective periods. 
In contrast $\{ banana \} {\rightarrow} \{ banana \}$ has a higher maximum intra-time ($\alpha_2 {=} 35$) and a lower average number of recurrences ($p {=} 14.63$). 
This indicates that: \emph{(i)} the time for a banana re-purchase is higher than the time of a milk re-purchase; \emph{(ii)} the support to have a distinct period is higher for $\{banana\}$ than $\{milk\}$.
We notice for more than 25\% of the customers the contemporary purchase $\{ bread, tomato \}$ can indicate a future basket with $\{ bovine \}$ or with $\{ banana, potato \}$ and that these TARS have very different annotations $\alpha, p, q$.
Finally, we highlight that, even if the most common TARS among the customers are those with base sequences, the TARS in $\Gamma_c$ with sequence length greater than two are on average more than the 95\% for each customer.

To better understand the TARS, in Table~\ref{tab:tars_coop} we shows some TARS made of base recurring sequences with different peculiarities.
A base recurring sequence capture the typical repurchasing of the same item within a certain period for a certain number of times.
\emph{Apples} and \emph{bananas} are fruit items available throughout they year. 
The associated base TARS $\{ banana \} {\rightarrow} \{ banana \}$ and $\{apple \} {\rightarrow} \{ apple \}$ have indeed a similar number of periods $p$ and number of typical occurrences in each period $q$. 
On the other hand, \emph{oranges} are a seasonal fruit item, generally available between November and February. 
The associated base TARS $\{ orange \} {\rightarrow} \{ orange \}$ has a recurrence $p$ significantly lower than the recurrence of banana and apple TARS, while the occurrence inside a period is similar. 
We observe that ice creams are similar to oranges: the associated TARS $\{ ice \; cream \} {\rightarrow} \{ ice \; cream \}$ has a lower $p$ and a higher maximum intra time $\alpha_2$.
Finally, \emph{Strawberries} and \emph{Easter eggs} are items available for just a short period of the year. 
As reflection in the associated TARS we have lower values of both $p$ and $q$ than the other TARS. 
In particular, among the items considered strawberries' TARS have the lowest $\alpha_2$ indicating short periods, while Easter eggs have the highest $\alpha_1$ indicating long intra-times.

\begin{table}[!tb]
\setlength{\tabcolsep}{0.01mm}
\centering
\begin{tabular}{cc}
\multicolumn{2}{l}{\ - \underline{Supported by more than 90\% customers}}\\[1mm]
\small $\{\mbox{milk}\} \xrightarrow[18.87, 6.58]{(1, 17)} \{\mbox{milk}\} $ & 
\small $ \{\mbox{banana}\} \xrightarrow[14.63, 7.20]{(2, 35)} \{\mbox{banana}\} $\\[3mm]

\multicolumn{2}{l}{\ - \underline{Supported by more than 80\% customers}}\\[1mm]
\small $\{\mbox{tomato}\} \xrightarrow[13.87, 6.58]{(1, 17)} \{\mbox{milk}\} $ & 
\small $ \{\mbox{tomato}\} \xrightarrow[15.27, 5.11]{(1, 12)} \{\mbox{bovine}\}$\\[3mm]
 
\multicolumn{2}{l}{\ - \underline{Supported by more than 25\% customers}}\\[1mm]
\small $ \bfrac{\mbox{bread},}{\mbox{potato}} \xrightarrow[11.40, 8.15]{[2, 15]} \{\mbox{bovine}\}$ & 
\small  $ \bfrac{\mbox{bread},}{\mbox{potato}} \xrightarrow[7.25, 4.30]{[3, 27]} \bfrac{\mbox{banana},}{\mbox{potato}}$\\[3mm]
\end{tabular}
\caption{Examples of TARS extracted from \textit{Coop-C}.}
\label{tab:tars_coop}
\end{table}

\begin{table}[!tb]
\hspace{-4mm}
\begin{minipage}[tb]{1.03\linewidth}
	\centering
	\includegraphics[width=\linewidth]{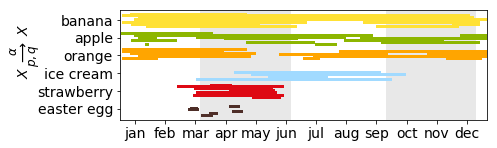}
\end{minipage}\hfill
\vspace{1mm}
\begin{minipage}[tb]{\linewidth}
	\centering
	\begin{tabular}{ccc|cccc}
	$X$ & $\rightarrow$ & $X$ & $\alpha_1$ & $\alpha_2$ & $p$ & $q$ \\
	\hline
	$\{ \mbox{banana} \}$ & $\rightarrow$ & $\{ \mbox{banana} \}$ & 2 & 35 & 14.63 & 7.20 \\
	$\{ \mbox{apple} \}$ & $\rightarrow$ & $\{ \mbox{apple} \}$ & 2 & 35 & 15.90 & 6.14 \\
	$\{ \mbox{orange} \}$ & $\rightarrow$ & $\{ \mbox{orange} \}$ & 2 & 33 & 8.13 & 6.56  \\
	$\{ \mbox{ice cream} \}$ & $\rightarrow$ & $\{ \mbox{ice cream} \}$ & 2 & 40 & 5.90 & 6.38 \\
	$\{ \mbox{strawberry} \}$ & $\rightarrow$ & $\{ \mbox{strawberry} \}$ & 2 & 32 & 3.55 & 4.69 \\
	$\{ \mbox{easter egg} \}$ & $\rightarrow$ & $\{ \mbox{easter egg} \}$ & 4 & 20 & 2.42 & 3.29
	\end{tabular}
\end{minipage}
\caption{TARS with different recurring base sequences from \textit{Coop-C}, and their periods shown along 7 years of observations (each year represented as a single line).}
\label{tab:tars_coop2}
\end{table}

\subsection{Properties of TBP}
In this section we present some peculiar properties of TBP: we show the temporal validity and reliability of the TARS extracted, and the performance improvements yield by the parameters evaluation.

\begin{figure}[tb]
    \centering
    	\hspace{-2mm}
   		\includegraphics[width=0.48\linewidth]{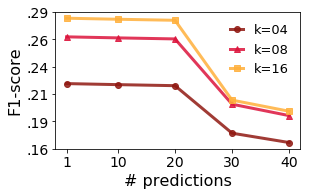}
   		\hspace{-2mm}
   		\includegraphics[width=0.48\linewidth]{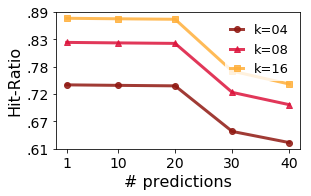}
\caption{Evaluation of TARS temporal validity on \emph{Coop-C}.}
\label{fig:exp_mp}
\end{figure}

\begin{figure}[tb]
    \centering
    	\hspace{-2mm}
   		\includegraphics[width=0.48\linewidth]{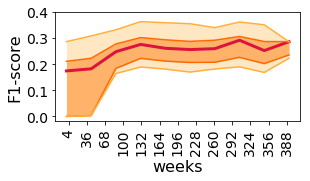}
   		\hspace{-2mm}
   		\includegraphics[width=0.48\linewidth]{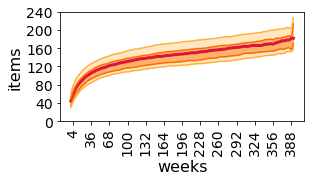}
   		
   		\hspace{-2mm}
   		\includegraphics[width=0.48\linewidth]{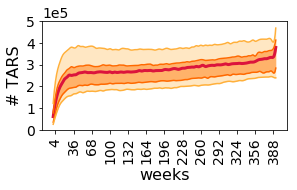}
   		\hspace{-2mm}
   		\includegraphics[width=0.48\linewidth]{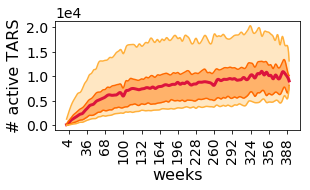}
   		
\caption{Evaluation of the TARS reliability by temporally augmenting the purchase history on \emph{Coop-C}.}
\label{fig:exp_growing_model}
\end{figure}

\subsubsection{TARS Temporal Validity}
\label{sec:exp_mult_pred}
In real-world applications it is unpractical (or even unnecessary) to rebuild a predictive model from scratch every time a new basket appears in a customer's purchase history.
This leads to the following question: for how long are TBP predictions reliable? 
We address this question by extracting TARS on the 70\% of the purchase history of every customer and performing the prediction on the subsequent baskets.
As shown in Figure~\ref{fig:exp_mp}, regardless the predicted basket size $k$, F1-score and Hit-Ratio remain stable up to $20$ predictions, which suggests a large temporal validity of TBP since the model construction. 

\subsubsection{TARS Extraction Reliability}
\label{sec:exp_model_rel}
How many baskets does TBP need to perform reliable predictions? 
For each customer we start from her second week of purchases and extract TARS incrementally by extending the training set one week at a time. We then predict the next basket of the customer and evaluate the performance of TBP in this scenario.
Figure~\ref{fig:exp_growing_model} shows the median value and the ``variance'' (by means of the 10th, 25th, 75th and 90th percentiles) of the F1-score (top-left), the total number of different items purchased by the customer (top-right), the number of TARS extracted (bottom-left), the number of active TARS during the prediction (bottom-right) as the number of weeks used in the learning phase increases.
On one hand the average F1-score does not change significantly as the number of weeks increases, while its ``variance'' reduces as more weeks are used in the learning phase. 
On the other hand, the other three measures stabilize after an initial setup phase. 

\subsubsection{Parameter-Free vs. Parameter-Fixed Approach}
\label{sec:exp_tars_eval}
\begin{figure}[tb]
    \centering
   		\includegraphics[width=0.48\linewidth]{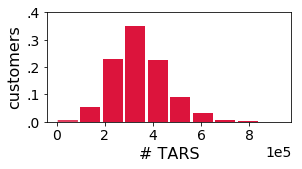}
   		\includegraphics[width=0.48\linewidth]{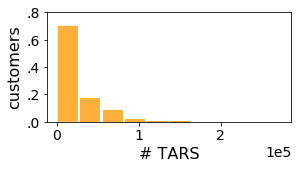}
\caption{Number of TARS per customer distribution on \emph{Coop-C}: parameter-free (left), parameter-fixed (right).}
\label{fig:exp_tars_eval1}
\end{figure}

TARS can be extracted by fixing the same parameters for all the customers and items, as usually done by state-of-the-art methods \cite{yu2016dynamic,wang2015learning,rendle2010factorizing,cumby2004predicting}, or by automatically estimating the parameters with a data driven procedure.
Here we discuss the impact of fixing the parameters on the predictive performance by comparing the results of parameter-free TBP and a parameter-fixed version of TBP where we set $\delta^{max}=14$ (e.g., two weeks), $q^{min} = 3$ and $p^{min} = 2$.
Figure~\ref{fig:exp_tars_eval1} shows the distributions of the number of TARS per customer for the parameter-free (left) and parameter-fixed (right) scenarios. 
We observe two different distributions: a skewed peaked distribution for the parameter-free scenario and a heavy tail distribution for the parameter-fixed scenario. 
This suggests that fixing the parameters has a strong impact on the extraction of TARS, leading to a lower average number of TARS per customer than the parameter-free scenario (Figure~\ref{fig:exp_tars_eval1}). 

\begin{figure}[tb]
    \centering
   		\includegraphics[width=0.48\linewidth]{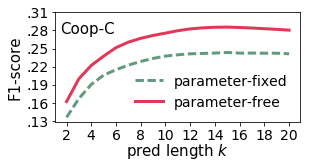}
   		\includegraphics[width=0.48\linewidth]{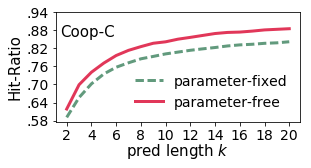}
\caption{Next basket prediction performance on \emph{Coop-C}: parameter-free (left), parameter-fixed (right).}
\label{fig:exp_tars_eval2}
\end{figure}

Figure~\ref{fig:exp_tars_eval2} compares the predictive performances of the parameter-free and the parameter-fixed scenarios. 
For both F1-score and Hit-Ratio, TBP produces better predictions in the parameter-free scenario. 
In particular, when using the average basket size of a customer $k^*_c$ as the size of the predicted basket, the parameter-free approach has F1-score=$0.25$ while the parameter-fixed approach has F1-score=$0.21$. 
Our results suggest that the adoption of a parameter-free strategy during the extraction of TARS enforces customer behavior heterogeneity and increases prediction accuracy.

\subsection{Comparison with Baseline Methods}
\label{sec:baselines}
We compare TBP with several baseline methods on the three datasets described above.
To this purpose, we implement\footnote{We provide the Python 2.7.11 source code of TBP and the baseline methods along with an anonymized sample of the \emph{Coop} dataset at this link https://goo.gl/JVsJcP.
The code of DRM was kindly provided by the authors of \cite{yu2016dynamic}.} the following user-centric state-of-the-art methods:

\textit{\textbf{LST}} \cite{cumby2004predicting}: the next basket predicted is just the \emph{last} basket purchased by the customer, i.e., $b_{t_{n+1}} = b_{t_{n}}$;

\textit{\textbf{TOP}} \cite{cumby2004predicting}: predicts the top-$k$ \emph{most frequent} items with respect to their appearance in the customer's purchase history $B_c$;

\textit{\textbf{MC}} \cite{cumby2004predicting}: makes the prediction based on the last purchase $b_{t_{n}}$ and on a \emph{Markov chain} calculated on the \mbox{customer's purchase history $B_c$;}

\textit{\textbf{CLF}} \cite{cumby2004predicting}: for each item $i$ purchased by the customer, this method builds a \emph{classifier} on temporal features extracted from the customer's purchase history considering two classes: ``item $i$ purchased yes/no''. The classifier then predicts the next basket by using the temporal features extracted from the customer's purchase history.

We also implement four state-of-the-art methods that are not user-centric, i.e, they require purchase data of all customers to build the predictive model for a single customer:

\textit{\textbf{NMF}} (\emph{Non-negative Matrix Factorization}) \cite{lee2001algorithms}: is a collaborative filtering method which applies a non-negative matrix factorization over the customers-items matrix. The matrix is constructed from the purchase history of all customers;

\textit{\textbf{FMC}} (\emph{Factorizing personalized Markov Chain})\cite{rendle2010factorizing}: 
combines Markov chains and 
matrix factorization to predict the next basket based on the purchase history of all the customers $\mathcal{B}$;

\textit{\textbf{HRM}} (\emph{Hierarchical Representation Model}) \cite{wang2015learning}: employs a two-layer structure to construct a hybrid representation over customers and items purchase history $\mathcal{B}$ from last transactions;

\textit{\textbf{DRM}} (\emph{Dynamic Recurrent basket Model}) \cite{yu2016dynamic}: 
it is based on recurrent neural network and 
captures both sequential features from all the baskets of a customer, and global sequential features from all the baskets of all the customers $\mathcal{B}$.

We compare TBP with the above defined baselines on \textit{Coop-A}, \textit{Coop-B} and \textit{Ta-Feng}. 
For NMF, FMC, HRM and DRM we report the results obtained with the default parameters setting \cite{yu2016dynamic,wang2015learning} and a dimensionality $d {=} 100$ for \emph{Ta-Feng}, \emph{Coop-A} and \emph{Coop-C}.
Even though the methods \cite{hsu2004mining,lazcorreta2008towards,wang2014modeling} employ patterns for producing recommendations we do not compare against them because they are systems mainly designed for web-based data services and because they are also consider the items' ratings and not only the implicit feedbacks provided by the presence of the items in a basket.

\begin{figure}[tb]
    \centering
    	\hspace{-1mm}
   		\includegraphics[width=0.48\linewidth]{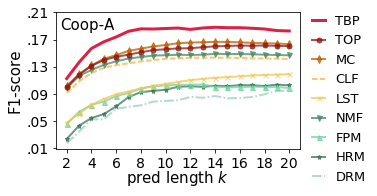}
   		\hspace{-10mm}
   		\includegraphics[width=0.48\linewidth]{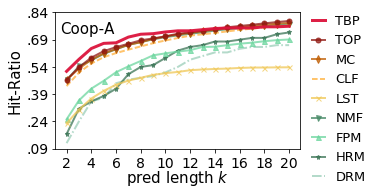}
		\vspace{-5mm}

   		\hspace{-1mm}
   		\includegraphics[width=0.48\linewidth]{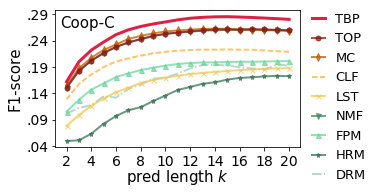}
   		\hspace{-10mm}
   		\includegraphics[width=0.48\linewidth]{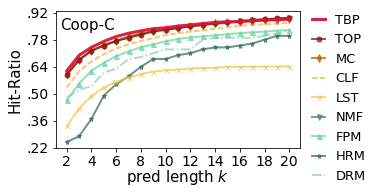}
   		\vspace{-5mm}

   		\hspace{-1mm}
   		\includegraphics[width=0.48\linewidth]{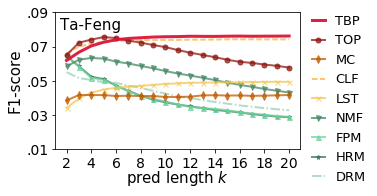}
   		\hspace{-10mm}
   		\includegraphics[width=0.48\linewidth]{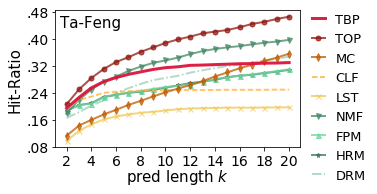}
   		\vspace{-7mm}

\caption{Performance comparison varying length $k$.}
\label{fig:exp_baseline1}
\end{figure}

Figure~\ref{fig:exp_baseline1} compares the average F1-score (left) and the average Hit-Ratio (right) produced by all the methods, varying $k \in [2, 20]$. 
We observe that TBP outperforms the competitors on \textit{Coop-A} and \textit{Coop-B}, having the highest F1-score and a comparable Hit-Ratio.
On \textit{Ta-Feng} TBP has the highest F1-score at the third highest Hit-Ratio. 
The performance of TBP significantly improves, both in terms of F1-score and Hit-Ratio, when we use $k = k^*_c$, as shown in Table~\ref{tab:exp_baseline2}.

The decrease of the Hit-Ratio of TBP in \textit{Ta-Feng} 
can be due to its 
high data sparsity. 
As we observe in Table~\ref{tab:dataset_stats}, \textit{Ta-Feng} 
has a much lower average number of baskets per customer, a much lower average basket length, and a shorter observation period than \emph{Coop-A} and \emph{Coop-C}. 
For this reason the TARS extracted from \textit{Ta-Feng} 
have lower quality than the TARS extracted on the other datasets.

\begin{table}[tb]
\setlength{\tabcolsep}{0.05mm}
\centering
\begin{tabular}{|c|c|ccccc|cccc|}
\hline
\multicolumn{2}{|c|}{$k=k_c^*$} & TBP &TOP & MC & CLF & LST & NMF & FPM & HRM & DRM \\
\hline
\rule{0pt}{2.5ex}
\multirow{3}{*}{\begin{turn}{90}$F1\mbox{-}score$\end{turn}}
 & \textit{Coop-A} & \textbf{.17} & \textit{\textbf{.14}} & \textit{\textbf{.14}} & .13 & .09 & \textit{\textbf{.14}} & .08 & .06 & .05 \\[0.5mm]
 & \textit{Coop-C} & \textbf{.24} & .22 & \textit{\textbf{.23}} & .19 & .14 & .22 & .16 & .08 & .12 \\[0.5mm]
 & \textit{Ta-Feng} & \textbf{.07} & \textbf{.07} & .04 & \textbf{.07} & .04 & \textit{\textbf{.06}} & \textit{\textbf{.06}} & \textit{\textbf{.06}} & .05 \\[0.5mm]
\hline
\rule{0pt}{2.5ex}
\multirow{3}{*}{\begin{turn}{90}$Hit\mbox{-}Ratio$\end{turn}}
 & \textit{Coop-A} & \textbf{.62} & .58 & .58 & .56 & .40 & \textit{\textbf{.59}} & .44 & .35 & .33 \\[0.5mm]
 & \textit{Coop-C} & \textbf{.72} & \textit{\textbf{.71}} & .70 & .65 & .50 & \textit{\textbf{.71}} & .61 & .38 & .55 \\[0.5mm]
 & \textit{Ta-Feng} & .20 & \textbf{.24} & .14 & \textit{\textbf{ .21}} & .15 & \textit{\textbf{ .21}} & \textit{\textbf{ .21}} & \textit{\textbf{ .21}} & .19 \\[0.5mm]
\hline
\end{tabular}
\caption{Performance using personalized length $k = k_c^*$. In \textbf{bold}, and \textit{\textbf{bold-italic}} the 1st and 2nd best performer.}
\label{tab:exp_baseline2}
\end{table}

In Table~\ref{tab:exp_buildtime} we report the duration of the learning process, i.e., the execution time needed to build every method on the four datasets. 
For the user-centric methods (TBP, MC, CLF) we report the average execution time per customer, while we report the total execution time for not user-centric methods (NMF, FPM, FRM, DRM).
We do not report the prediction time because it is negligible for all the approaches (i.e., less than 0.01 seconds).
We observe that TBP needs more time than existing user-centric methods (5 minutes per customer on average) but it is much faster than the not user-centric approaches. 
We believe that such a learning time is acceptable for two reasons: \emph{(i)} in a real scenario the TARS can be re-computed once every month and still produce reliable predictions; \emph{(ii)} the computation can be parallelized and personalized with respect to the customer's behavior, thus the TARS of all the customers can be extracted at the same time by different devices.

\begin{table}[tb]
\setlength{\tabcolsep}{0.05mm}
\centering
\begin{tabular}{|c|rrr|rrrr|}
\hline
Dataset & \multicolumn{1}{c}{TBP} & \multicolumn{1}{c}{MC} & \multicolumn{1}{c|}{CLF} & \multicolumn{1}{c}{NMF} & \multicolumn{1}{c}{FPM} & \multicolumn{1}{c}{HRM$^*$} & \multicolumn{1}{c|}{DRM$^*$} \\
\hline
\textit{Coop-A} & 351.86s  & 0.04s & 2.38s & 244.28s & 0.21h & 0.84h & 47.53h \\
\textit{Coop-C} & 6.62s & 0.01s & 1.08s & 69.98s & 0.11h & 0.72h & 34.06h \\
\textit{Ta-Feng} & 0.01s & 0.00s & 0.00s & 803.89s & 0.41h & 0.34h & 4.24h \\
\hline
\end{tabular}
\caption{Building time comparison. Note that: \emph{(i)} the time is reported in seconds (s) or in hours (h), \emph{(ii)} for individual methods we report the average building time, for collective methods the total, \emph{(iii)} the building time for TOP and LST is always lower than $0.01$ seconds and is not reported.}
\label{tab:exp_buildtime}
\end{table}

\begin{figure}[tb]
    \centering
    	\hspace{-1mm}
   		\includegraphics[width=0.48\linewidth]{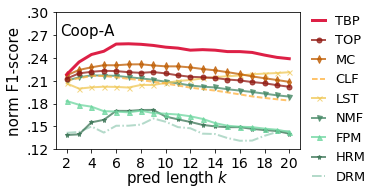}
   		\hspace{-10mm}
   		\includegraphics[width=0.48\linewidth]{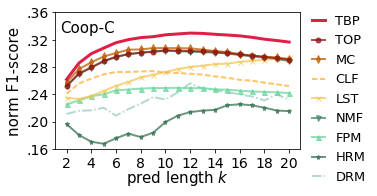}
\caption{Normalized F1-score varying length $k$.}
\label{fig:exp_baseline3}
\end{figure}

It is worth noting that the value of the average F1-score can be biased by two extreme scenarios: \emph{(i)} 
the F1-score can be low because of a low Hit-Ratio, i.e., for most of the customers no item is predicted even though for some customers we predict most of the items; \emph{(ii)} 
the F1-score can be high because for most of the customers just one item is predicted. 
In Figure~\ref{fig:exp_baseline3} we show the results of the experiments using the normalized F1-score instead of the simple F1-score. We observe that the positive gap between TBP and the competitors increases: for the customers for which TBP correctly predicts at least one future basket, the baskets predicted by TBP are more accurate and cover a larger number of items than the baskets predicted by the other methods.

\begin{figure}[tb]
    \centering
    	\hspace{-1mm}
   		\includegraphics[width=0.48\linewidth]{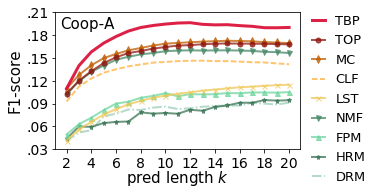}
   		\hspace{-10mm}
   		\includegraphics[width=0.48\linewidth]{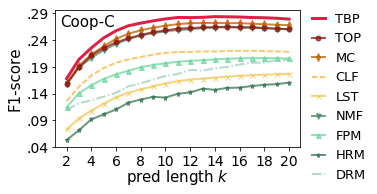}
		\vspace{-8mm}
\caption{Performance comparison varying the prediction length $k$ and using a model built on a subset of $B_c$ having random size between 70\% and 90\% of $|B_c|$.}
\label{fig:exp_baseline4}
\end{figure}

We also investigate at what extent the performances can be affected by the \emph{leave-one-out} evaluation strategy: the last basket of a customer could depart from her typical behavior affecting the extraction of the TARS. To cope with this issue we also perform the learning process (i.e., extract TARS) by selecting a random subset ${B'}_c = \{ b_{t_1}, \dots, b_{t_{n'}} \}$ of the customers' purchase history $B_c = \{ b_{t_1}, \dots, b_{t_n}\}$, with $t_{n'} < t_{n}$. 
We randomly vary the size of the subset $|{B'}_c|$ among 70\% and 90\% of $|B_c|$, and we apply TBP on the subsequent basket $b_{t_{n'+1}}$. 
Figure~\ref{fig:exp_baseline4} presents the results of this experiment for \textit{Coop-A} and \textit{Coop-C} and confirms the trends observed on the previous experiments indicating that the \emph{leave-one-out} evaluation strategy does not affect significantly \mbox{the performance of the methods.}

\section{Conclusions}
\label{sec:conclusion}
In this work we propose a data-driven and user-centric approach for next basket prediction.
Our contribution is twofold.
First we define Temporal Annotated Recurring Sequences (TARS).
Then we then use TARS to build a predictor for forecasting customers' next baskets. 
Being parameter-free, TBP leverages the specificity of individual customer's behavior to adjust the way TARS are extracted, thus producing more personalized patterns. 
We perform experiments on real-world datasets, show that TBP outperforms state-of-the-art methods and, in contrast with them, it provides interpretable patterns that can be used to gather insights on customers' shopping behaviors. 
Our results show that at least 36 weeks of a customer's purchase behavior are needed to effectively predict her next baskets. 
In this scenario, TBP can effectively predict the subsequent twenty future baskets 
with remarkable accuracy.

A future research line consists in providing to the customers of a living laboratory \cite{vescovi2014my} an app running TBP and observe how and if their purchase behaviors are influenced by the recommendations.
Moreover, since our method is fully user-centric, it cannot make reliable predictions for new customers or for customers having a short purchase history. 
Thus, we plan to build a version of TBP which incorporates a collaborative filtering approach, such that it will be able to forecast baskets for newcomers and for customer with a short purchase history.
Finally, we would like to exploit TARS also to segment the customers, and to investigate how TARS and TBP can be applied on different analytical domains such as for mobility and for health data analysis.


{
\small
This work is partially supported by the GS501100001809 European Communitys H2020 Program under the funding scheme ``INFRAIA-1-2014-2015: Research Infrastructures''
grant agreement, http://www.sobigdata.eu, GS501100001809, 654024 \emph{``SoBigData: Social Mining \& Big Data Ecosystem''}.

}

\bibliographystyle{abbrv}
\bibliography{biblio} 

\end{document}